%% file: main.tex
\documentclass[conference]{IEEEtran}
\IEEEoverridecommandlockouts
\usepackage{cite}
\usepackage{amsmath,amssymb,amsfonts}
\usepackage{algorithmic}
\usepackage{graphicx}
\usepackage{textcomp}
\usepackage{xcolor}
\usepackage{booktabs} 
\usepackage{float}

\usepackage{longtable}
\usepackage{array}
\usepackage{placeins}

\usepackage{ulem}
\usepackage{xcolor}
\newcommand{\changed}[1]{\textcolor{black}{#1}} 

\def\BibTeX{{\rm B\kern-.05em{\sc i\kern-.025em b}\kern-.08em
    T\kern-.1667em\lower.7ex\hbox{E}\kern-.125emX}}

\makeatletter
\newcommand{\linebreakand}{%
  \end{@IEEEauthorhalign}
  \hfill\mbox{}\par
  \mbox{}\hfill\begin{@IEEEauthorhalign}
}
\makeatother
\begin{document}

\title{Students' Perceptions of the Use of LLMs in Requirements Engineering Education: A Cross-University Empirical Study\\
}

\author{\IEEEauthorblockN{1\textsuperscript{st} Sharon Guardado}
\IEEEauthorblockA{\textit{M3S Research Unit} \\
\textit{University of Oulu}\\
Oulu, Finland \\
sharon.guardadomedina@oulu.fi}
\and
\IEEEauthorblockN{2\textsuperscript{nd} Risha Parveen}
\IEEEauthorblockA{
\textit{Tampere University}\\
Tampere, Finland \\
risha.parveen@tuni.fi}
\and
\IEEEauthorblockN{3\textsuperscript{rd} Zheying Zhang}
\IEEEauthorblockA{ \textit{Tampere University}\\
Tampere, Finland \\
zheying.zhang@tuni.fi}
\linebreakand
\IEEEauthorblockN{4\textsuperscript{th}Maruf Rayhan }
\IEEEauthorblockA{
\textit{Tampere University}\\
Tampere, Finland\\
maruf.rayhan@tuni.fi}
\and
\
\IEEEauthorblockN{5\textsuperscript{th} Nirnaya Tripathi}
\IEEEauthorblockA{\textit{M3S Research Unit} \\
\textit{University of Oulu}\\
Oulu, Finland\\
nirnaya.tripathi@oulu.fi}
}

\maketitle

\begin{abstract}
The integration of Large Language Models (LLMs) in Requirements Engineering (RE) education is reshaping pedagogical approaches, seeking to enhance student engagement and motivation while providing practical tools to support their professional future. This study 
empirically evaluates the impact of integrating LLMs in RE coursework. We examined how the guided use of LLMs 
influenced students' learning experiences,  
and what benefits and challenges they perceived in using LLMs in RE practices. The study collected survey data from 179 students across two RE courses in two universities. LLMs were integrated into coursework through different instructional formats, i.e. individual assignments versus a team-based Agile project.  
Our findings indicate that LLMs improved students’ comprehension of RE concepts, particularly in tasks like requirements elicitation and documentation. 
However, students raised concerns about LLMs in education, including academic integrity, overreliance on AI, and challenges in integrating AI-generated content into assignments. Students who worked on individual assignments perceived that they benefited more than those who worked on team-based assignments, highlighting the importance of contextual AI integration. 
This study offers recommendations for the effective integration of LLMs in RE education. It proposes future research directions for balancing AI-assisted learning with critical thinking and collaborative practices in RE courses. 

\end{abstract}

\begin{IEEEkeywords}
Large Language Models,
Requirements Engineering,
AI-assisted Learning,
Higher Education,
Students perspective,
Empirical study,
Cross-institutional research
\end{IEEEkeywords}

\input{sections/introduction}

\input{sections/background}

\input{sections/context}

\input{sections/researchdesign}

\input{sections/result}
\input{sections/discussion}
\input{sections/conclusion}

\bibliographystyle{IEEEtran}
\bibliography{references}

\end{document}

%% file: sections/introduction.tex
\section{Introduction}

Requirements Engineering (RE) plays a crucial role in software development by ensuring that systems meet the needs of stakeholders and business objectives\cite{wiegers2013software}. As RE education (REE) evolves to meet the demands of an increasingly complex software landscape, innovative teaching approaches are necessary to equip students with both theoretical knowledge and practical skills. An emerging approach is the integration of Large Language Models (LLMs) into REE, which offers opportunities to improve the learning experience, increase student participation, and provide intelligent support for RE tasks\cite{alier2025lamb}\cite{zhang2024llm}.
LLMs such as ChatGPT have gained prominence in educational settings, particularly in software engineering disciplines\cite{lo2023impact}\cite{hou2024}. These AI-driven tools can assist in RE tasks such as eliciting, documenting, and validating requirements. 
However, the incorporation of LLMs into REE raises important questions regarding their effectiveness, ethical implications, and impact on student learning outcomes. While LLMs can potentially enhance student understanding of concepts, concerns such as overreliance on AI-generated content and the accuracy of their outputs must be carefully considered \cite{zhai2024}.

\changed{Although the integration of LLMs into education is gaining momentum, there remains a lack of robust empirical evidence on how to implement them effectively, particularly from the viewpoint of students enrolled in RE courses\cite{chan2023students}. To our knowledge, only two previous studies have explored the guided integration of LLMs in REE. Both were single-institution pilot interventions involving fewer than 50 students, and focused narrowly on specific tasks such as requirements documentation and refinement \cite{sampaio2024exploring}\cite{moravanszky2024banning}. This highlights a significant empirical gap in understanding the broader pedagogical value of LLMs throughout the entire lifecycle of RE.}

This study \changed{addresses the identified gap with the first cross-institutional study investigating the guided use of LLMs in REE. We analyse the use of LLMs to support RE courses across two universities, Tampere University (UA) and University of Oulu (UB). Both universities provide courses designed to enhance students’ grasp of RE concepts and practices. UA does so by integrating individual assignments with group projects, while UB adopts a project-based, hands-on approach within an Agile software development (ASD) context.} 

Our study gathered data from 179 students enrolled in the RE courses, where LLMs were integrated into the coursework in a guided manner. In UA, LLMs were used in individual assignments, whereas in UB, the LLMs were used in team-based assignments. In both cases, we analysed \changed{the students' perceived effectiveness of LLMs in facilitating key RE tasks, including project planning, requirements elicitation, specification, and prioritisation. This allowed us to explore how the guided use of LLMs influenced students' understanding of RE concepts, their course engagement, their perceptions of the benefits and challenges associated with AI-assisted learning, and their overall motivation to learn.} 
The following research questions guide the research:
\begin{itemize}
    \item \textbf{RQ1}: How do students perceive the impact of integrating LLMs on their RE learning experience?
    \item \textbf{RQ2}: What are students’ perceived benefits and challenges of LLM-supported RE practices?
\end{itemize}

Our research aims to advance the knowledge of AI-supported REE by assessing the advantages and limitations of purposely guiding the integration of LLMs into RE courses. By understanding the impact of AI on student learning, educators can better design instructional strategies that leverage the strengths of LLMs while adequately addressing the challenges they pose for education.

%% file: sections/background.tex
\section{Background}

\subsection{Requirements Engineering Education}

RE is a fundamental discipline in software engineering, which contributes to the development of software systems that meet the needs of stakeholders and align with business goals \cite{sommerville1997requirements}\cite{wiegers2013software}\cite{tripathi2018anatomy}. Effective RE practices are essential to mitigate the risks of project failure, budget overruns, and misaligned or incomplete system functionalities \cite{cheng2007research}. Additionally, RE is inherently socially intensive, which demands smooth interactions with various stakeholders to reach a consensus on the specifications of a software system. 
In line with this, REE faces unique challenges in preparing future engineers to manage the complexities of RE practices. REE must not only provide students with theoretical knowledge and techniques but also develop their ability to handle sociotechnical aspects, including effective stakeholder communication, negotiation, and conflict resolution. However, simulating real-world social interactions in a classroom remains a significant hurdle. Without sufficient guidance and extended instructor involvement \cite{ouhbi2015requirements, neves2021teacher, ALHARBI2024ENH}, it is often difficult for students to experience the real-world social interactions inherent to RE processes, which are critical for developing the necessary competencies they will require in their future RE practices.  
REE employs diverse teaching methods and instructional approaches to address these challenges. The traditional lecture-based methods have been complemented by experiential learning techniques such as case-based learning, project-based learning, and role-playing exercises \cite{zowghi2003teaching}\cite{regev2009experiential}\cite{svensson2017role}. These methods aim to bridge the gap between theory and practice, allowing students to participate in simulated RE scenarios \cite{davis2006communication}. Many reports on educational experiences have highlighted the benefits of incorporating stakeholder roles in the coursework to improve students' understanding of RE practices. However, these reports also underline the challenges in replicating the complexity of real-world scenarios within classroom settings, aligning academic instruction with evolving industry practices, and fostering essential soft skills such as communication, collaboration, and critical thinking \cite{cheng2007research}. 

Addressing these challenges requires innovative pedagogical approaches. Integrating AI-driven tools, such as LLMs, provides an opportunity to enhance REE by simulating realistic stakeholder roles \cite{zhang2024llm} and interactions while providing adaptive feedback to students. Although traditional REE methods have contributed to foundational knowledge and skills, the evolving landscape of software engineering demands more innovative, AI-driven approaches to address persistent educational challenges and form the focus of this study.

\subsection{The use of LLMs in REE}
LLMs are AI models designed to understand, generate, and process human language.  They use generative capabilities to produce text-based outputs, including essays, summaries, translations, code, and conversations \cite{zhang2021ai}. Tools such as ChatGPT, Bard and AI-powered tutoring systems provide instant feedback and explanations, while predictive analytics assist in identifying at-risk students, facilitating timely intervention \cite{lo2023impact}. 

When utilised in software engineering education, LLMs have been shown to convey multiple benefits, such as enhancing learning experiences by helping students with coding assignments, debugging, and concept explanations, as evidenced by the articles on AI-related topics\cite{seet2024}\cite{rasnayaka2024llms}. In addition, LLMs can support self-paced learning, allowing students to work on coding problems at their own pace\cite{Silva2024}.
Likewise, researchers have investigated how educators can utilise LLMs to improve other instructional activities, such as automating feedback provision \cite{balse2023evaluating} and optimising the grading process \cite{wieser2023investigating}. These advancements highlight the potential of  LLMs to streamline teaching practices and improve student learning experiences in software engineering and its sub-disciplines. Although students generally report having positive experiences with LLMs, instructors often express mixed reactions. Some of the most common concerns are that LLMs may reduce students' problem-solving efforts and that there is a negative correlation between LLM use and grades, in which students who rely heavily on LLMs tend to struggle with basic tasks\cite{Raihan2025}. Prior research on the use of LLMs in software engineering education and its subfields emphasises the importance of preserving students' critical thinking and problem-solving abilities. In the context of RRE, LLMs should be positioned as tools that enhance cognitive engagement. Their integration must prioritise learning enrichment, ensuring they serve as a supportive aid rather than a substitute for the development of core engineering competencies\cite{khan2025llms}.

%% file: sections/context.tex
\section{Research context}


With the growing emphasis on RE in software engineering curricula and the increasing role of LLMs in education, it is essential to explore how these technologies can be effectively integrated into RE learning. 
To address this gap, this study explores the guided integration of LLMs into RE courses offered at UA and UB. These universities were chosen based on their well-established software engineering programs, each offering REE with distinct pedagogical approaches. UA follows a blended learning approach with individual assignments and group projects, while UB adopts a project-based, hands-on learning model in the context of ASD. Despite these differences, the RE courses offered in both universities share the common objective of improving the understanding of RE concepts and practices by integrating LLMs into the coursework. To ensure comparability, similar RE tasks and consistent instructions on the use of LLMs in assignment completion were provided in both courses. \changed{Although the study is confined to two institutions, their approaches to REE are consistent with widely recognised pedagogical practices in software engineering programs in higher education. Consequently, the findings retain strong relevance and potential applicability to comparable academic contexts.}

\subsection{Course delivered by UA}


The Requirements Engineering (REQ) course is a 5 ECTS\footnote{ECTS stands for European Credit Transfer and Accumulation System. It is used to transfer and accumulate credits between higher education institutions. One ECTS credit equals 25-30 hours of effort.} credits course in the Master’s Degree Programme in Computing Sciences and Electrical Engineering at UA, primarily designed for students in software engineering. 
It is offered annually in the Autumn semester, with approximately 120 participants each year. 
The course highlights the critical role of RE in software development, system analysis, and product management and covers both theoretical foundations and practices in requirements elicitation, analysis, specification, validation, and requirements management across diverse software projects. Before taking the course, students should have basic knowledge of software engineering. 
The course uses a blended learning approach, consisting of 20 hours of lectures, 10 weekly assignments, a mastery exercise, and group work. All learning materials, including lecture notes, additional readings, assignments, and project guidelines, are shared through the university Moodle platform. Students complete individual assignments based on their understanding of lectures and textbooks. 
After six weeks of lectures, students start collaboratively working on a project in teams of three or four, selecting a predefined or self-proposed topic. The project allows students to apply RE knowledge in a research or practice-oriented context. In addition, a mastery exercise, in the form of an online quiz, assesses students' grasp of the RE concepts. 

Although assignments, group work, and mastery exercises are not compulsory, active participation is essential to impact the final grade. This structured yet flexible format equips students with analytical, critical thinking, and collaborative skills and prepares them to address complex challenges in modern software development.

\subsection{Course delivered by UB} 

At UB, the Professional Software Engineering and Human Factors (PSEHF) course is also 5 ECTS credits and has been designed for bachelor's and master’s degree programs in software engineering and information systems. It has been offered in the Autumn semester for the past four years (2021-2024), with 70-90 participants annually. The grade is evenly divided, with 50\% allocated to individual work and 50\% to the group project. The project is a key component of the course and follows a project-based, hands-on learning model where Agile practices are used, guiding students through an iterative, team-based ASD process.
Throughout the course, students collaborate in teams of four or five on a predefined project. They function as a Scrum team, applying Agile practices with industry-supported tools. Some of their responsibilities include creating user stories, managing a product backlog, and developing a working application prototype. The course instructors act as customers throughout multiple one-week sprints. Jira is utilised as a project management tool, where requirements are collected and refined, and where the project progress is documented. Each team selects their preferred communication platform. Most teams utilise Discord, WhatsApp or MS Teams. Scrum ceremonies, including sprint planning, daily stand-ups, sprint reviews, and sprint retrospectives, are mandatory to plan, capture, document, and prioritise the requirements for each sprint. Documentation is part of the deliverables expected each week. By the end of the project, students gain hands-on experience in RE using ASD practices and tools.

\subsection{Integrating LLMs in coursework}

\begin{table*} [t]  
\small  
\setlength{\tabcolsep}{4pt} 
\centering  
\caption{Tasks and Learning Objectives}
\label{tab:assignments}
\scriptsize
\begin{tabular}{|p{1.35cm}|p{2cm}|p{4.5cm}|p{1.5cm}|p{4.5cm}|p{2.5cm}|}
    \hline
    \multicolumn{1}{|c|}{\textbf{Task}} & 
    \multicolumn{1}{|c|}{\textbf{Learning objective}} & 
    \multicolumn{2}{|c|}{\textbf{REQ course}}& 
    \multicolumn{2}{|c|}{\textbf{PSEHF course}}\\
    \hline
    & & \textbf{Task description}& \textbf{Deliverables}&  \textbf{Task description}&\textbf{Deliverables}\\
    \hline
    \textbf{Project planning}& 
    Formulate a clear project vision and stakeholder analysis & 
    1. Draft a vision statement, perform a Need, Approach, Benefits, Competition analysis (NABC), and identify stakeholders. \newline
    2. Use an LLM to refine and validate the analysis and description in Step 1. \newline
    3. Discuss the AI's impact in project planning based on Steps 1 \& 2.& 
    Project vision \& scope document& 
     1. Identify stakeholders and conduct a NABC analysis focusing
on their needs. \newline 2. Use ChatGPT to explore why the stakeholders and NABC analysis are useful and how they are conducted.& Project plan \& stakeholders analysis   \\
    \hline
    \textbf{Requirements elicitation}& 
    Apply appropriate techniques for requirements elicitation & 
    Role-playing to elicit requirements using an LLM simulating stakeholders with an elicitation technique like interviews, brainstorming, etc. & 
        Requirements description & 
    Without   ChatGPT, elicit functional and non-functional requirements. Document the requirements with user stories.\newline& Functional and non-functional requirements as user stories\\
    \hline
    \textbf{Requirements specification}& 
    Enhance clarity, consistency, and completeness of requirements & 
    1. Manual improvement of requirements. \newline
    2. Enhancement using an LLM. \newline
    3. Compare AI-enhanced vs. manually improved versions. & 
        Refined requirements& 
     1. Select user stories that need
improvement. Use ChatGPT  for enhancement. Evaluate if
its suggestions are valid. \newline 2. Using the ChatGPT, simulate an elicitation technique with one of your potential users.& Analysis of user stories enhancement with  LLMs. List of new requirements obtained with the elicitation technique.\\
    \hline
    \textbf{Requirements prioritisation}& 
    Analyse the value and cost of requirements and prioritize them using a proper technique & 
    1. Analyse and prioritise requirements using a selected technique. \newline
    2. Use an LLM for the same prioritisation. \newline
    3. Compare AI-generated vs. student-generated results. & 
    Prioritised requirements & 
    Use ChatGPT to research prioritisation techniques and effort estimation. Evaluate the pros and cons of different techniques and select one to prioritize the user stories in your backlog.& User stories estimation and backlog prioritisation.\\
    \hline
    
\end{tabular}
\label{tab:tasks}
\end{table*}

The guided use of LLMs in coursework aimed to enhance students' learning experiences, encourage critical thinking, and simulate real-world RE scenarios. The coursework was designed to align with key learning objectives while guiding students to explore the benefits and limitations of LLMs in RE practices.
The assignment formats differed between the two courses due to variations in learning objectives, content, and implementation. However, both courses shared common RE tasks, including project planning, requirements elicitation, prioritisation, and specification. These shared tasks provided the foundation for examining the impact of a guided use of LLMs across different instructional approaches. A sample of the learning objectives and task descriptions of both courses is shown in Table \ref{tab:assignments}. 
Although the REQ course integrated LLMs into individual assignments, and the PSEHF course embedded them into team-based project assignments, both courses maintained a consistent approach regarding guidelines for the use of LLMs. Before working on the first assignment, students received guidance on using LLMs in the course, including practical tips for effectively crafting prompts.

The REQ course adopted a structured three-step approach for each assignment:

\begin{itemize}
    \item Manual completion – Students independently complete a given task, such as stakeholder identification, requirements elicitation, etc.
    \item LLMs integration - After completing the task, students used an LLM of their choice to refine, validate, or replicate their work.
    \item Analysis and reflection – Students documented their interactions with the LLM, compared AI-generated results with the human-led work, and reflected on how LLMs contributed to or influenced their decision-making in task completion.
\end{itemize}

This approach encouraged students to critically evaluate the reliability and usefulness of AI-generated content.

The PSEHF course adopted a similar approach by integrating LLMs into the ASD project. At the beginning of each sprint, students were provided with detailed task instructions, including deliverable templates and explicit guidelines on where and when to use LLMs and when they should not be used. Students were first required to complete the task without LLMs, and in a subsequent stage, they could utilise LLMs for the same task, but they were required to analyse the results and decide whether to use the LLM-generated result. \changed{As an example, for the requirements elicitation and refinement tasks, students were instructed to proceed as follows: 
\begin{itemize}
    \item  Using the project plan and the NABC, identify your platform's initial list of requirements from the end-users' point of view. The suggested user story template is "As a [type of user], I want [goal] so that [reason/benefit]. (Do not use LLMs for this task.)
    \item After eliciting the requirement, select a few of them that could be improved. Use ChatGPT to suggest enhancements to the chosen requirements.
    \item  Analyse whether the suggested improvements are valid and decide whether or not to implement them. Briefly document the differences between your own outputs and those generated by ChatGPT.
\end{itemize}}
\changed{Table \ref{tab:samples} provides real examples of students' own analysis on their interactions with LLMs to refine the requirements.}

\begin{table*}[t]
\scriptsize  
\setlength{\tabcolsep}{4pt} 
\centering  
\changed{ \caption{Students' Sample Analyses on Improving Requirements Using LLMs}}
    \begin{tabular}{|p{0.8cm}|p{2.8cm}|p{5.7cm}|p{4.3cm}|p{3cm}|}\hline
        \parbox[c]{0.8cm}{\centering \textbf{Course}} & 
\parbox[c]{2.8cm}{\centering \textbf{Original requirement}} & 
\parbox[c]{5.7cm}{\centering \textbf{LLM-enhanced requirement}} & 
\parbox[c]{4.3cm}{\centering \textbf{How did the LLM enhanced the requirement}?} & 
\parbox[c]{3cm}{\centering \textbf{Usefulness of LLM-provided suggestion}} \\\hline
         
         \textbf{REQ} &  As a user I want the tracker to monitor my health data in real-time so that I can keep track of my well-being continuously & As a user, I want the fitness tracker to monitor and display real-time health metrics (heart rate, blood pressure, calories, stress) so I can track my health during exercise. Includes acceptance criteria: live heart rate display, alerts if thresholds exceeded, and updates within five seconds. & LLM added clarity by writing the requirement as a user story, adding acceptance criteria, making it testable, and specifying thresholds and timeframes, which were missing in my manual version. & Using ChatGPT was much faster and gave clearer output. I would have preferred starting with it. \\\hline
         \textbf{PSEHF}&  As an event planner, I want to communicate with providers in the platform&  As an event planner, I want to communicate seamlessly with providers through the platform to ensure efficient collaboration and quick responses. In addition to feature suggestions like integrated video calls, it also suggested things like conflict resolution with a report button, cross-device synchronization, and many others.& ChatGPT increased the descriptive power of the user story. It included many things we had not even considered. A lot of the ideas are very useful.& We decided not to implement some of the ideas, because they do not provide the core functionality that is required of an Minimal Viable Product.\\\hline
    \end{tabular}
   \label{tab:samples}
\end{table*}

In each assignment, specific recommendations for the effective use of LLMs were provided in both courses. For example, for the task of prioritising requirements, students were advised to: \textit{"Use an LLM to research techniques for backlog prioritisation and effort estimation. Prompt the LLM to evaluate the pros and cons of the different techniques and to guide you on how to implement the selected techniques".} By integrating these structured guidelines, students were able to experiment with LLMs as learning aids, maintaining a balance between independent problem-solving and AI assistance.
 
The guided use of LLMs throughout different assignments provided a controlled yet flexible framework for students to explore the role of LLMs in RE practices, while critical reflection components ensured that LLMs were used as learning enhancers rather than shortcuts for assignment completion.

%% file: sections/researchdesign.tex
\section{Research design}


This study investigates the impact of the guided use of LLMs in RE courses in two universities. Assignments were central in both courses, helping students understand theoretical concepts and develop critical thinking and problem-solving skills. By incorporating LLMs as interactive peer learning tools in coursework, we aimed to enhance their learning experience, encourage idea exploration and provide timely guidance.
Our study objectives are twofold: first, to examine students’ perceptions of the impact of LLMs on their RE learning process, and second, to identify the perceived benefits and challenges associated with the use of LLMs in RE practices. Additionally, we explore how LLMs contribute to innovative teaching methods and student engagement. To achieve these objectives, we conducted a voluntary survey at the end of both courses. It aimed to evaluate the effectiveness of LLMs in supporting RE practices and enriching students’ overall learning experience. The target population was students participating in the RE courses. The following section describes the study design, data collection, and analysis process.

\subsection{Questionnaire design}

To design the study questionnaire, we followed a multi-stage design process \cite{dillman2014internet}. This process began with clarifying the study objectives and identifying the target population, which was aligned with the specific learning outcomes of the RE courses. \changed{Simultaneously, we reviewed prior literature on the use of LLMs in educational settings, particularly examining the most commonly reported benefits and challenges \cite{amoozadeh2024trust}\cite{marques2024using}.} The questionnaire was developed in accordance with established guidelines for survey instrument design \cite{fowler2013survey}\cite{dillman2014internet}, ensuring clarity and ease of comprehension. Items were grouped thematically and presented on separate pages with clear headings to facilitate understanding. A cover page outlined the survey’s purpose, emphasising the importance of participation to advance AI integration in REE. It also detailed the data usage and storage procedures, assuring respondents of voluntary and anonymous participation, and providing contact details for inquiries. 
The first section of the questionnaire collected demographic information, including students’ current level of studies and prior experience using LLMs. The second section examined students’ experience using LLMs in the coursework, focusing on study motivation, engagement, comprehension of RE concepts, and assignment completion. The third section explored students’ perceptions of the benefits and challenges of integrating LLMs in RE practices, covering key RE tasks embedded in coursework, i.e. stakeholder analysis, requirements elicitation, prioritisation, and documentation.
The course instructor at UA created the initial questionnaire draft, which was reviewed and further refined in collaboration with course instructors and teaching assistants at UA and UB. To ensure the questionnaire was contextually appropriate for each university, terminology was adapted as needed. For instance,  while the REQ course allowed the use of any LLM,  the PSEHF course primarily focused on the use of ChatGPT. Consequently, references to “LLM” in the questionnaire were replaced with “ChatGPT” where appropriate. After reaching a consensus, the final version of the questionnaire was made available online using Microsoft Office Forms. It comprised 17 mandatory closed-ended questions, including Likert scale and multiple-choice items, as well as four optional open-ended questions that allowed students to further elaborate on their opinions and experiences. To validate the clarity and usability of the questionnaire, the final version was pilot-tested by two teaching assistants, one from each university. A copy of the questionnaire and a summary of the results are available here\footnote{https://doi.org/10.6084/m9.figshare.29210438.v1}.
\subsection{Data Collection}
The survey link was made available on the Moodle page of each course. The REQ course was conducted from September 4 to December 6, 2025, with a total of 117 students. The survey was launched on October 9, coinciding with the final week of the individual assignments, and remained open for two weeks. The PSEHF course ran from October 30 to December 23, with 98 participants. In this case, the survey was made available on November 18 and remained open for three weeks.
By the time they completed the survey, students in both courses had engaged with the guided use of LLMs for a minimum of one month, ensuring they had sufficient exposure to provide informed responses. 
As survey participation was entirely voluntary, both courses implemented incentives to encourage responses. 
The REQ course offered 2.5 bonus points, while the PSEHF course offered two bonus points for completing the survey.  
To ensure anonymity, 
students at UA could submit their student ID separately. 
At UB, students had to complete the survey to receive instructions for marking the activity as completed in Moodle. Importantly, the course grading structure was designed to ensure that students who chose not to participate would not be penalised or disadvantaged in their final evaluations.

At the end of the data collection period, the average completion time for the survey was 15 minutes and 10 seconds. UA received 105 responses, while UB received 78. However, for the sake of consistency, four responses from UB were excluded as they were submitted by students from different academic levels (e.g., Ph.D. and Open University). This resulted in 74 responses from UB, bringing the total number of survey responses for the study to 179. 

\subsection{Data Analysis}

The data analysis began by exploring potential relationships between key variables. Spearman’s rank correlation and Chi-Square test were used to assess associations between variables such as prior experience with LLMs, study level, and students’ perceptions of various aspects of LLM use. While Spearman’s correlation coefficients ranged from weak to moderate, none of the relationships reached statistical significance, suggesting no meaningful associations among the variables analysed. Given this, we proceeded to analyse the data through comparative analysis to evaluate students' perceptions and further investigate patterns in the dataset. Descriptive statistics were used to summarise responses to closed-ended questions, including frequency distributions and percentage agreement levels. To assess potential differences between universities, 
we examined response distributions and compared agreement levels across different categories.
Although our primary focus was on quantitative trends, we reviewed qualitative responses from the open-ended survey items to provide a contextual understanding of the data and compare the results with the descriptive statistics analysis. Relevant student quotes were extracted to support the findings of the comparative analysis.

%% file: sections/result.tex
\section{Results}
A total of 179 students participated in the survey, the majority of whom (83.8\%) were enrolled in a master's degree program. At the beginning of the questionnaire, students were asked to self-assess their prior experience with LLMs. Over half  (53.1\%) identified as intermediate users, while more than one-third (34.1\%) considered themselves beginners. Demographic information of the participants is presented in Table \ref{tab:demographics}. Most participants reported prior use of LLMs in other courses, primarily for concept understanding (147/179), code documentation and explanation (126/179), and language improvement or translation (118/179).

\begin{table}[t]  
\small  
\setlength{\tabcolsep}{4pt} 
\centering  
\caption{Participant Demographics by University}
\label{tab:demographics}
\begin{tabular}{|l|l|l|l|}
    \hline
    \multicolumn{1}{|c|}{\textbf{Demographics}} & 
    \multicolumn{1}{c|}{\textbf{REQ Course}} & 
    \multicolumn{1}{c|}{\textbf{PSEHF course}} & 
    \multicolumn{1}{c|}{\textbf{Total}} \\
    \hline
    \multicolumn{4}{|c|}{\textbf{Study level}} \\
    \hline
    Master's & 91 & 59 & 150 \\
    Undergraduate & 14& 15& 29\\
    \hline
    \multicolumn{4}{|c|}{\textbf{Prior experience with LLMs}} \\
    \hline
    No experience & 5 & 0 & 5 \\
    Beginner & 44 & 17 & 61 \\
    Intermediate & 51 & 44 & 95 \\
    Advanced & 5 & 13 & 18 \\
     \hline
    \multicolumn{1}{|c|}{\textbf{Total}} & 
    \multicolumn{1}{c|}{\textbf{105}} & 
    \multicolumn{1}{c|}{\textbf{74}} & 
    \multicolumn{1}{c|}{\textbf{179}} \\
    \hline
\end{tabular}
\end{table}

\subsection{RQ1: How do students perceive the impact of integrating LLMs on their RE learning experience?}
To assess the effect of a guided integration of LLMs on students' learning experiences, we conducted a quantitative analysis of multiple closed-ended survey questions, complemented by a qualitative analysis of open-ended responses.  
The survey items related to RQ1 explored how LLMs contributed to understanding RE concepts, how they supported assignment completion, the perceived benefits and challenges of using LLMs, and how LLMs influenced students' motivation and overall learning process.
\subsubsection{\textbf{Enhanced understanding of RE concepts and requirements analysis}}
\begin{figure}[h]
    \centering
    \includegraphics[width=1\linewidth]{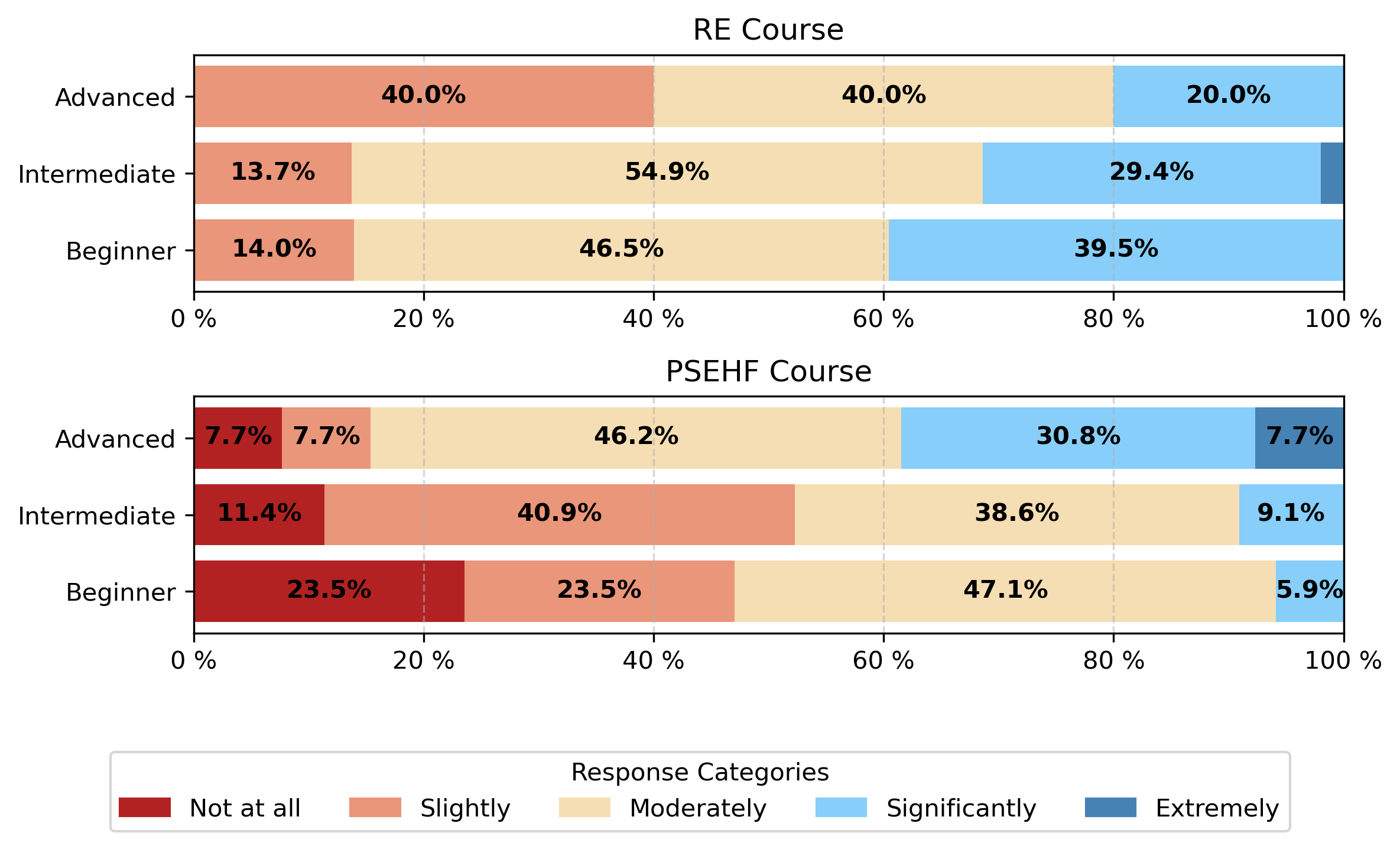}
    \caption{Impact of LLMs on understanding RE concepts}
    \label{fig:enhancement}
\end{figure}

Overall, students reported a positive perception of LLMs in supporting their learning. More than 70\% of the participants indicated that LLMs helped to improve their understanding of RE concepts. Students with advanced experience were more likely to recognise this benefit in both courses. As illustrated in Fig.\ref{fig:enhancement}, respondents in the REQ course were more prone to perceive LLMs as beneficial for understanding RE concepts than their counterparts in the PSEHF course. In the PSEHF course, 10 of the 74 respondents stated that LLMs did not contribute to enhancing their understanding of RE concepts. This perception was more prevalent among beginners (23.5\%). 

A similar trend was observed in students' perceptions of LLMs' effectiveness in enhancing their ability to analyse requirements. Among advanced users, an overwhelming 94.1\% acknowledged this benefit of LLMs. Across all levels of expertise, 74.3\% expressed a positive perception, while beginners reported a slightly lower agreement rate of 68.9\%.
These findings suggest that prior experience plays a role in students' confidence in using LLMs to analyse requirements, with more experienced users deriving greater benefits.  


\subsubsection{\textbf{Influence on motivation and engagement in learning}}
\begin{figure}[h]
    \centering
    \includegraphics[width=1\linewidth]{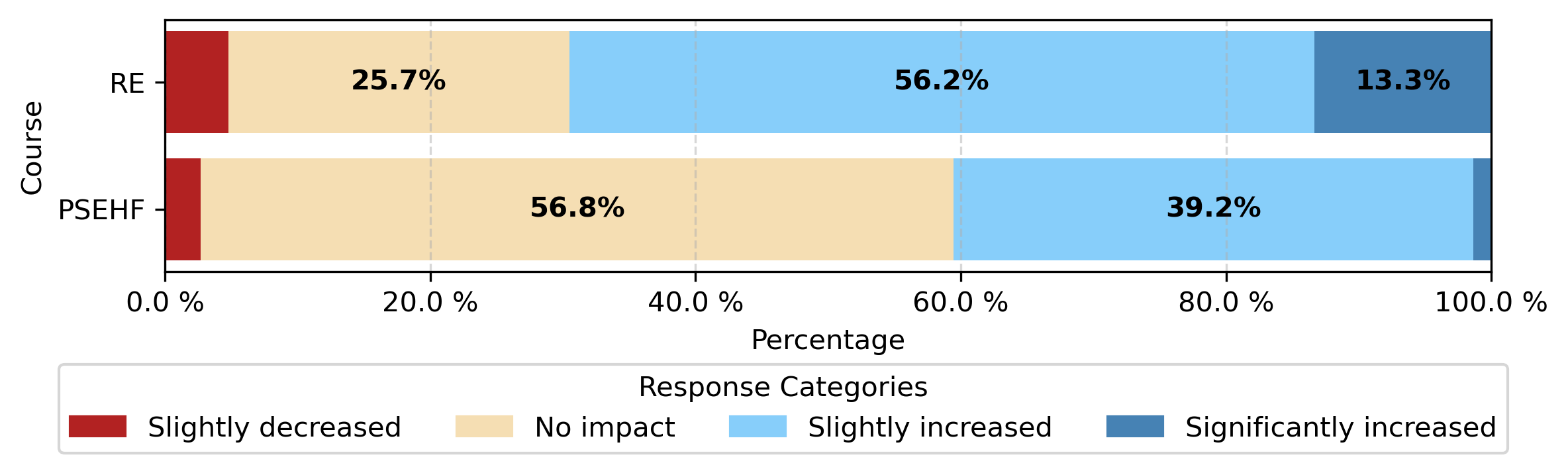}
    \caption{Influence of LLMs on student's motivation to learn RE}
    \label{fig:motivation}
\end{figure}

In both courses, the integration of LLMs had a positive effect on students' motivation to learn. As shown in Fig.\ref{fig:motivation}, 
This effect was more significant in the REQ course, where nearly 70\% reported a slight to significant increase in their motivation. On the contrary, in the PSEHF course, more than half of the students (56.8\%) stated that using LLMs did not impact their motivation. 
Students' responses highlight that the guided use of LLMs positively influenced their motivation to learn RE. No students reported a significant decrease in motivation, and only a very small percentage (3.9\%) indicated experiencing a slight decrease. This suggests that using LLMs in the course did not hinder motivation in most cases.

Furthermore, 111 out of 179 students indicated that using LLMs made learning more engaging. This perception remained consistent across universities, different levels of education and previous experience, pointing to the potential of LLMs to foster a more engaging learning environment regardless of the students' familiarity with the technology. 

\subsubsection{\textbf{Perception of the integration of LLMs in coursework}}
\begin{figure}
    \centering
    \includegraphics[width=1\linewidth]{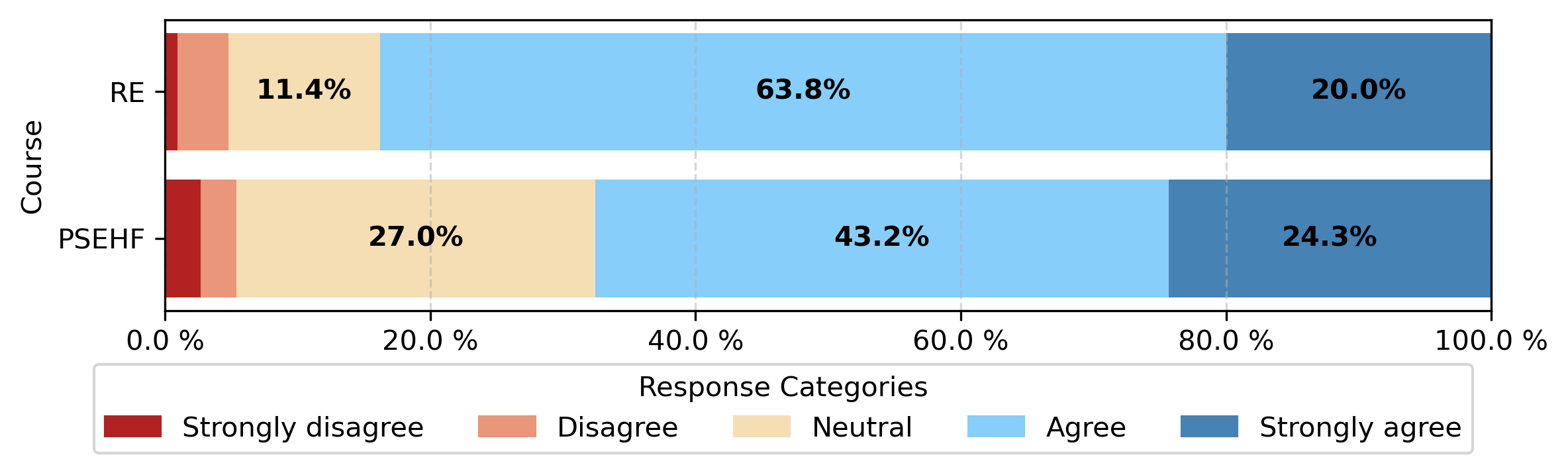}
    \caption{Agreement with the integration of LLMs being smooth and effective}
    \label{fig:smoothinte}
\end{figure}

Overall, the majority of students perceived the guided integration of LLMs as smooth and effective. Only 5\% of the respondents 
disagreed (Fig. \ref{fig:smoothinte}).
A key benefit widely identified by students was the efficiency gained through LLM-assisted learning, particularly in reducing the time required to complete tasks. Advanced users especially acknowledged this advantage, with 17 out of 18 agreeing.
Although some challenges were reported, they did not significantly overshadow the students' overall experience. 
Notably, two-thirds of students in both courses (66.5\%) support the use of LLMs in future courses. Among respondents who were opposed to integrating LLMs in future courses, concerns primarily centred on ethical implications and academic integrity. As one respondent (P71) noted: \textit{“LLMs should not be a mandatory part of the course, and their use should not be encouraged because of ethical concerns}”.

 In UB, where students had to work as a Scrum team, 12 out of 17 who perceived themselves as beginners indicated that using LLMs was not helpful or only slightly helpful in completing the weekly assignments. Their perception could be influenced by the additional challenges faced in trying to work collaboratively within an ASD approach. About this challenge, one respondent (P77) commented: \textit{"It's important to recognise that real Agile RE requires active stakeholder collaboration and team dynamics that AI cannot fully replicate; some of the challenges of using LLMs relate to its limited ability to understand changing project dynamics and team relationships".}

\subsubsection{\textbf{Challenges in using LLMs during the course}}
\begin{figure}
    \centering
    \includegraphics[width=1\linewidth]{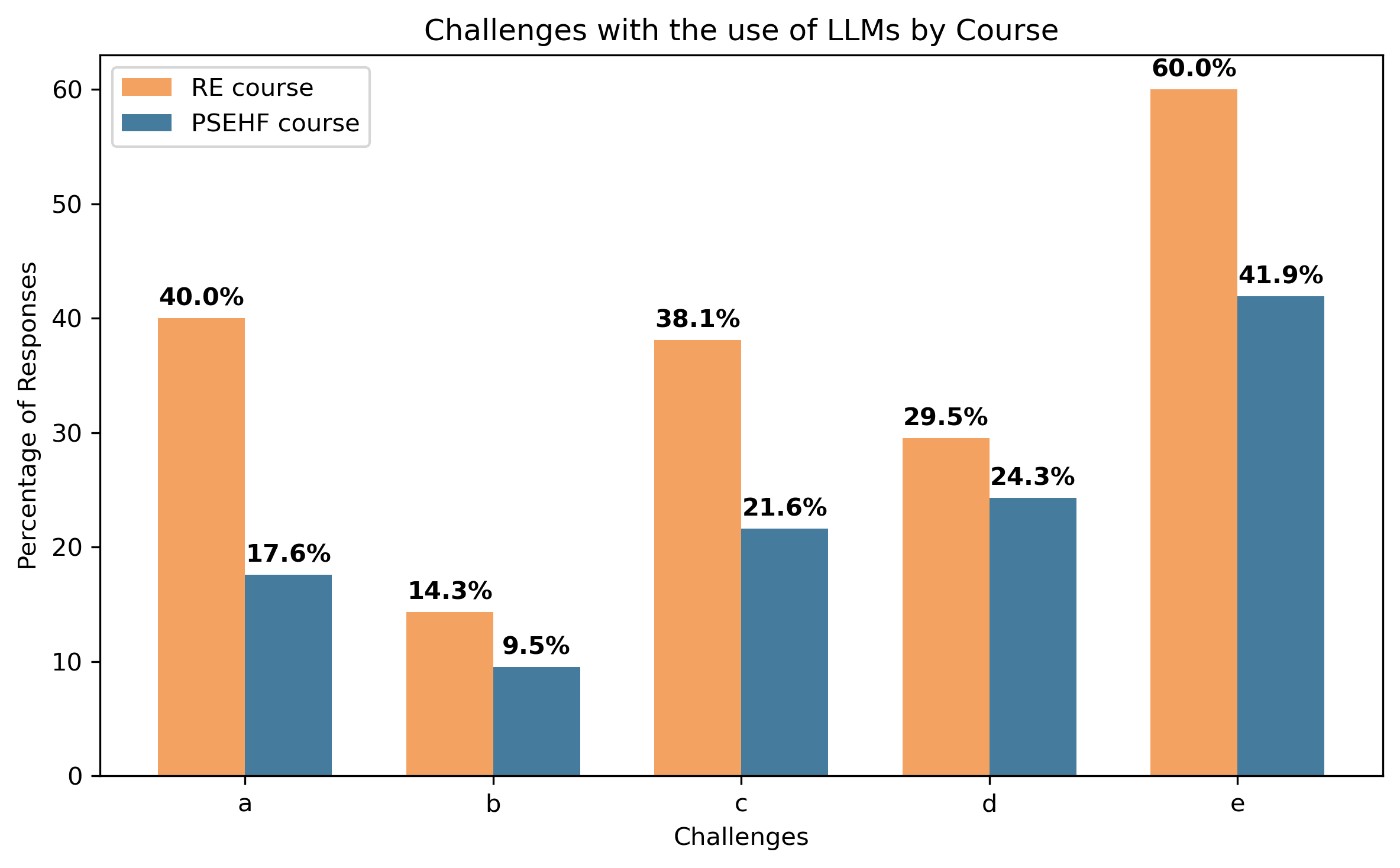}
    \caption{\textbf{The most common challenges students perceived in using LLMs in the course.}
    Labels: 
    \textbf{a.} Understanding how to use LLMs effectively, 
    \textbf{b.} Technical issues (access or usability), 
    \textbf{c.} Overreliance on LLMs for answers, 
    \textbf{d.} Difficulty integrating LLM-generated outputs, 
    \textbf{e.} Concern about academic integrity.
  }
    \label{fig:challuni}
\end{figure}

140 out of 179 students reported facing at least one challenge when using LLMs. As illustrated in Fig.\ref{fig:challuni}, the most frequently cited challenges were concerns about academic integrity, i.e. using LLMs ethically and responsibly in the coursework. Many students struggled with the ethical implications of appropriately balancing AI-generated content with their own effort, or how to reference LLMs in their work correctly. 
One student (P108) noted: \textit{"It concerned me that I was putting an excessive amount of effort into writing personal answers if, to some extent, I could get away with using AI-generated responses."}
Other challenges included difficulties interpreting LLM-generated outputs or uncertainty about providing sufficient context to ensure accurate responses. Although general guidelines and tips for effectively crafting prompts were shared with students throughout the courses, 17.6\% of students in the PSEHF course and 40\% of students in the REQ course found it challenging to write effective prompts. Novice LLMs users struggled the most. As illustrated by a respondent (P80): \textit{“As a beginner user of LLMs, I'm not that competent at 'prompt engineering' of how to converse with LLMs effectively”}.

These results suggest that while ethical concerns remain an essential discussion point, the practical benefits of LLMs in supporting coursework are widely appreciated. However, students require more explicit guidelines on ethical and responsible AI use, best practices for structuring queries, and strategies for critically assessing AI-generated content. 


\subsection{RQ2: What are students’ perceived benefits and challenges of using LLM-supported RE practices?}

 
 
 
 

To investigate students’ perspectives on the benefits and challenges of using LLMs in RE practices, we analysed survey responses focusing on their experiences with LLM-assisted RE tasks, perceived advantages, and concerns. Our analysis highlights how LLMs contributed to creativity, productivity, and efficiency in RE while revealing concerns about accuracy, overreliance, and ethical considerations. By examining the answers to closed-ended and open-ended questions, we gained insight into students' attitudes toward adopting LLMs in RE and the factors that influence their trust and engagement with AI-driven tools.

\begin{figure*}[t]
    \centering
    \includegraphics[width=0.73\textwidth]{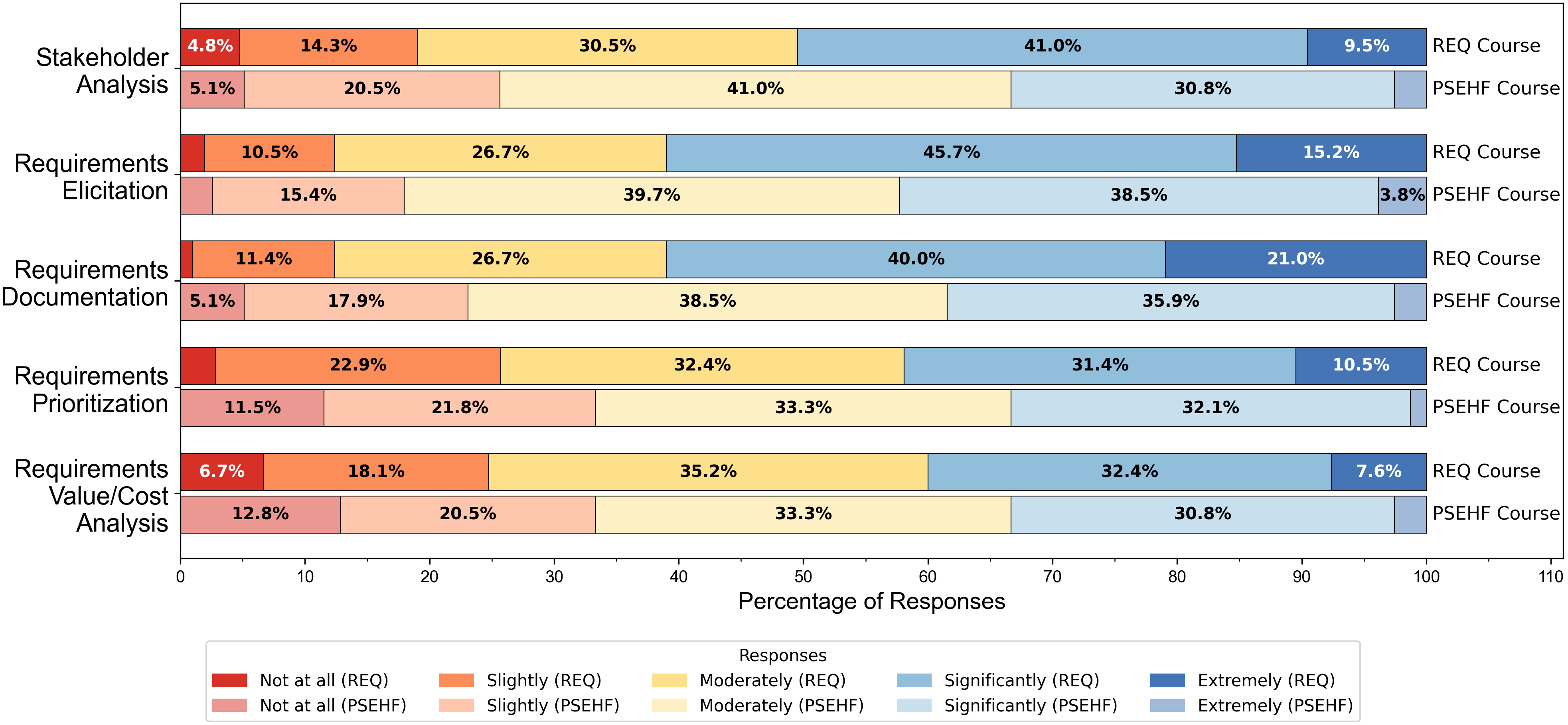}
    \caption{ Students' perception of the support of LLMs for RE tasks}
    \label{fig:q12-perceived-benefits}
\end{figure*}

\subsubsection{\textbf{Perceived benefits from using LLMs in RE}}
Students identified the most significant benefits of using LLMs for requirements elicitation and documentation (Fig. \ref{fig:q12-perceived-benefits}). In the REQ course, more than 60\% of the students found LLMs significantly or highly beneficial for these tasks, making them the most positively rated. Although students in the PSEHF course exhibited a similar trend, their perceptions were generally more moderate. This difference may be influenced by the coursework structure, i.e. individual assignments versus team-based projects. In individual work, LLMs served as the primary peer assistants with whom the students interacted intensively. In contrast, team-based coursework requires discussions and collaboration with other students, potentially reducing the need to rely on LLMs. A student in the REQ course (P18) noted, “\textit{Using LLMs was especially helpful in making a comprehensive stakeholder analysis, as well as for refining and clarifying requirements}”. This view was echoed by 18 of the 71 respondents. In contrast, value/cost analysis of requirements was perceived as the task for which LLMs were least beneficial. However, about 40\% of REQ course students and a third of PSEHF students still found LLMs significantly helpful in this task.
\begin{figure}[h!]
    \centering
    \includegraphics[width=1\linewidth]{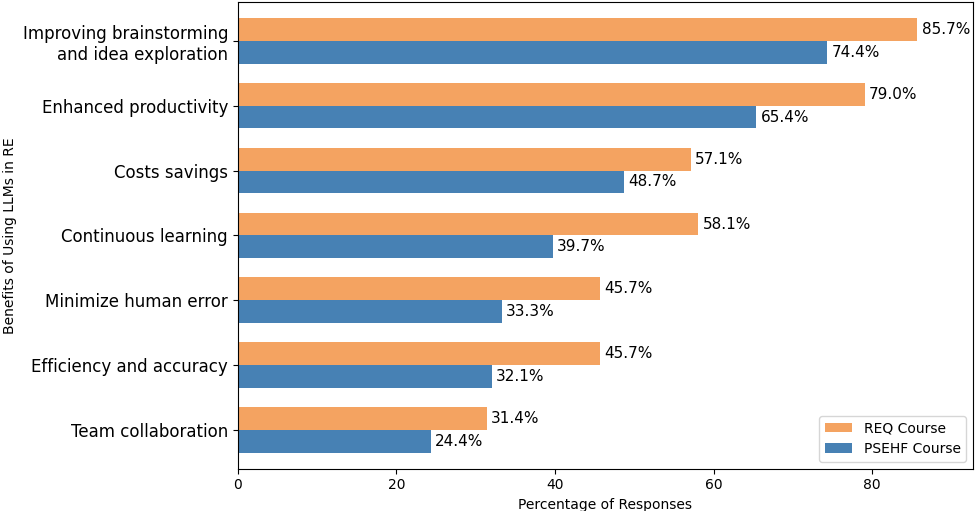}
    \caption{Students' perceived benefits of using LLMs in future RE practices}
    \label{fig:q16-broader-benefits}
\end{figure}

Fig. \ref{fig:q16-broader-benefits} presents the perceived benefits of using LLMs in their future RE practices, as reported by students. The response patterns were consistent across universities, with the most frequently endorsed advantage being "improving brainstorming and idea exploration" and "enhanced productivity". Students in the REQ course reported slightly higher agreement, with 85.7\% and 79.0\% endorsing these benefits, respectively. This confirms students' acknowledgement of the role of LLMs in stimulating creativity and boosting efficiency.
As one student (P99) noted, 
“\textit{LLMs always provided more viewpoints than I figured out myself".} This sentiment was shared by 28 out of 71 respondents.
Similarly, many students emphasised the support of LLMs for time savings on repetitive tasks, allowing more focus on complex RE problem-solving. A student (P79) with prior Scrum Master experience stated, “\textit{A lot of developer time is spent on writing specifications and requirements, refining stories and epics. I believe that using LLMs for refinement, especially, could have a positive impact on the work}”. 
LLMs were also perceived to be beneficial for cost savings, continuous learning, and minimising human error. 
While LLMs were perceived as useful for individual tasks, their impact on teamwork was unclear. Team collaboration was the least endorsed benefit of using LLMs, aligning with our findings from the previous section. 
Overall, these findings highlight a strong consensus from students on the ability of LLMs to enhance brainstorming, productivity, and cost-effectiveness while also shedding light on areas where their impact is weaker, particularly in teamwork.

\subsubsection{\textbf{Challenges and Concerns}}

\begin{figure*}
    \centering
    \includegraphics[width=0.73\textwidth]{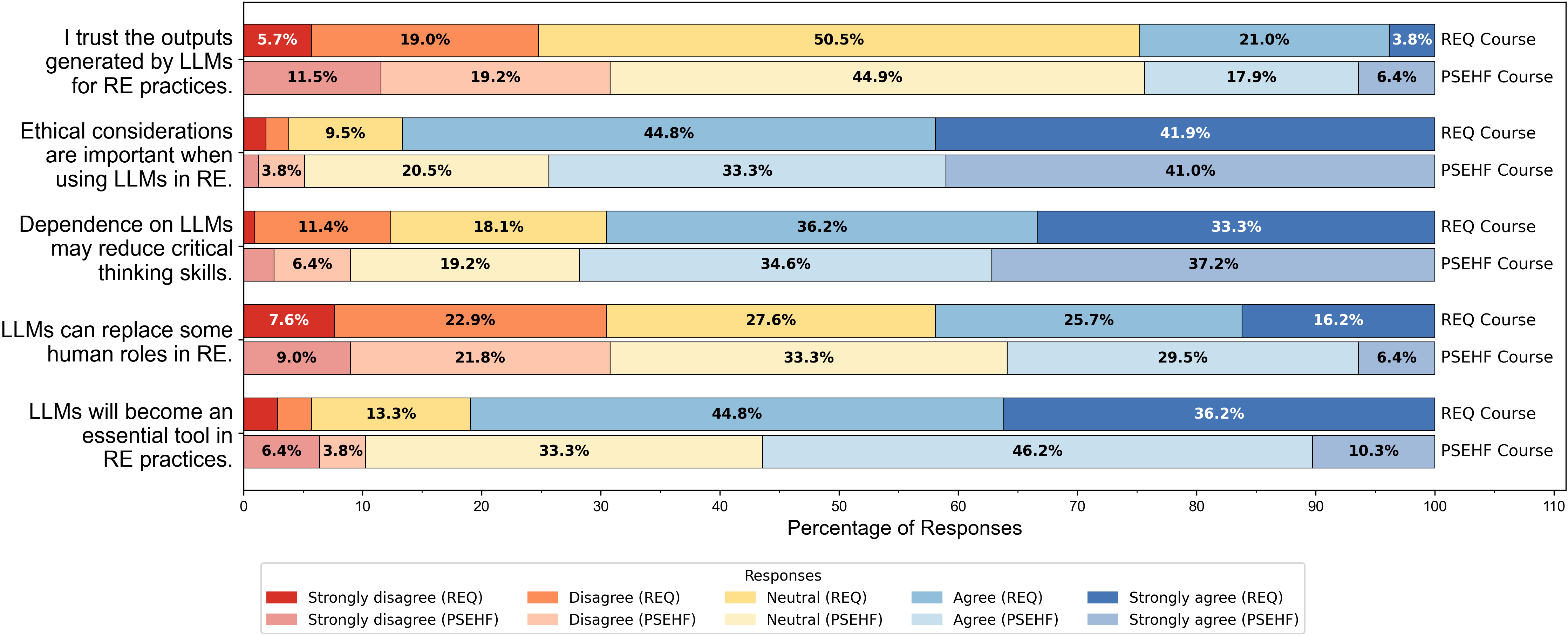}
    \caption{Students' level of agreement with statements regarding the use of LLMs in RE}
    \label{fig:q14-attitude}
\end{figure*}

\begin{figure}
    \centering
    \includegraphics[width=1\linewidth]{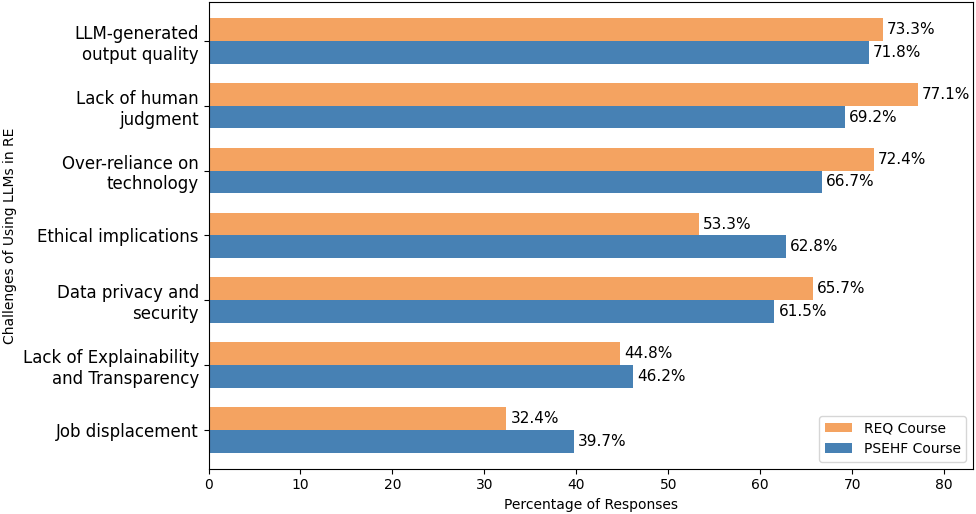}
    \caption{Students' perceived challenges of using LLMs in future RE practices}
    \label{fig:q15-key-concerns}
\end{figure}

As illustrated in Fig. \ref{fig:q14-attitude}, many students now view LLMs as essential tools. 81\% of the students in the REQ course agreed or strongly agreed with this view, compared to 56.5\% among PSEHF students. Despite this recognition, students are still cautious about AI-generated content. Over 50\% of REQ students and 44.9\% of PSEHF students remained neutral about their trust in LLMs, and over 25\% expressed some level of distrust. This scepticism aligns with the key challenges shown in Fig. \ref{fig:q15-key-concerns}, where LLM-generated output quality was the most cited challenge, with more than 70\% of respondents in both courses raising concerns about accuracy, consistency, and hallucinations. As one student (P34) remarked, “\textit{Sometimes, I am not sure about the accuracy of the LLM-generated outputs, so I have to research more on the Internet and the lecture slides}”. Students in the PSEHF course exhibited greater scepticism, emphasising critical evaluation before relying on LLM-generated content. A student (P10) reinforced this sentiment: “\textit{Never trust ChatGPT’s outputs. Always read them and ensure that they correctly satisfy the assignment}”. 

Similarly, students expressed concerns about bias, fairness, and accountability. More than 70\% of students were concerned about future overreliance on AI (Fig. \ref{fig:q15-key-concerns}). In their current studies, respondents were also concerned about weakened critical thinking skills due to dependence on LLMs (Fig. \ref{fig:q14-attitude}). Many students acknowledged that while LLMs can support them in RE tasks, excessive dependence could hinder analytical and problem-solving abilities. As a student (P9) reflected: “\textit{To me, usage of ChatGPT should be moderate. Otherwise, it will kill my thinking and creativity}”. Additionally, 77.1\% of those in the REQ course and 69.2\% of those in the PSEHF course expressed doubts about the ability of LLMs to make context-aware decisions, emphasising the "lack of human judgment" challenge in Fig. \ref{fig:q15-key-concerns}. This assertion was further supported by the concerns on explainability and transparency, with about half of the students questioning how LLMs generate responses and whether their reasoning could be trusted. While the students at UA more strongly agreed that "ethical considerations are important when using LLMs in RE" (Fig. \ref{fig:q14-attitude}), students at UB were more apprehensive about practical ethical implications, such as academic integrity and responsible AI use (Fig. \ref{fig:q15-key-concerns}). Students recognised the importance of both ethical guidelines and the real-world ethical challenges when integrating LLMs into teamwork. 
Data privacy and security risks were another major concern, with over 60\% of students expressing worries about handling sensitive information. As one student (P38) pointed out, “\textit{If LLMs are not hosted by the company itself, sensitive data can be leaked to third parties}”. 

Overall, some students showed mixed feelings about using LLMs in RE tasks and in their future RE practices. While they recognised their potential to improve efficiency, creativity, and productivity in RE, they remained cautious about reliability, ethical risks, and loss of critical thinking skills.  

%% file: sections/discussion.tex
\section{Discussion}
\subsection{Comparative analysis}
The distribution of students' responses was similar across both universities, with only minor variations attributable to factors such as study level or perceived prior experience. However, notable differences emerged in a few specific areas. These differences concerned how students perceived the extent to which LLMs enhanced their understanding of RE concepts, the support they provided for the completion of weekly assignments, and how they contributed to the students' learning engagement.
In these three areas, students from UA demonstrated more consistent responses than those from UB. This suggests that UA students, regardless of their prior experience with LLMs, held a more uniformly positive perception of the guided use of LLMs in the course.
 

Overall, the integration of LLMs proved more beneficial for students in the RE course, where assignments were completed individually. In contrast, students in the PSEHF course encountered additional challenges due to the need to self-organise and collaborate within a Scrum team while operating under a tight project timeline.

Throughout the one-month project, PSEHF students were required not only to grasp RE concepts and complete the associated assignments but also to learn and apply the Scrum framework, participate in regular Scrum ceremonies, become familiar with project management tools, and develop an application prototype. These added layers of complexity likely increased their cognitive workload and time pressure, which may have decreased the perceived usefulness of LLMs to support task completion.


For most topics, there were no significant differences in perceptions between undergraduate and master's students. However, notable variations emerged based on students' prior expertise with LLMs. This finding aligns with previous research done in multiple disciplines, where it was found that students with prior exposure to generative AI tools reported more positive perceptions regarding their utility in learning tasks\cite{chan2023students}. Students with limited experience using LLMs are less likely to perceive their benefits, even within guided learning contexts, so there is a clear need to develop targeted training and resources. These resources should be designed to support novice users in effectively leveraging LLMs to enhance their learning and assist with different RE tasks.
\subsection{Benefits and Challenges of Using LLMs in REE}
Our findings suggest that course structure and team dynamics play a critical role in shaping students’ perceptions of LLMs within REE. Students from UA, who worked individually, reported greater perceived benefits from using LLMs in tasks such as brainstorming, stakeholder analysis, and documentation, where they could freely utilise and refine AI-generated content.
In contrast, students from UB, who operated in ASD teams, expressed more scepticism regarding the guided use of LLMs in the course. This hesitancy may stem from the need to incorporate detailed project context when crafting prompts to reduce inaccuracies in AI-generated outputs. Furthermore, the collaborative nature of the team-based work, where multiple members reviewed and contributed to each task, may have amplified concerns about accuracy and the potential over-reliance on LLMs by certain team members. A limitation of current LLMs is their lack of direct support for collaborative work. While they can assist individuals in streamlining specific tasks, they do not inherently facilitate group decision-making processes. This was echoed in feedback from several UB students, who emphasised that RE relies heavily on negotiation and human judgment that AI, at present, cannot fully replicate.


Ethical and data privacy concerns also followed this pattern. UA students viewed privacy as an institutional issue, whereas UB students working with a real-world project were more concerned about data security risks when integrating AI into their workflows. This suggests that exposure to practical AI challenges increases awareness of associated ethical and privacy challenges.
These findings highlight the need for context-specific AI training in REE. For individual learners, emphasis should be placed on mitigating the risk of overreliance on LLM-generated content. For learners working in teams, guidance is needed to support the responsible and effective integration of AI into collaborative, real-time workflows. 

\subsection{Comparison to related work and our study contribution}

\changed{Existing studies on the use of LLMs in REE, such as \cite{carvallo2023use} \cite{sampaio2024exploring} \cite{mellqvistexploring} \cite{abdelfattah2023roadmap} highlight their potential to improve pedagogical approaches, but vary in both scope and depth. Carvallo and Erazo-Garzón \cite{carvallo2023use} demonstrated the value of LLMs in project-based learning to understand concepts and simulate interviews. However, they noted issues with contextual accuracy. The authors in \cite{sampaio2024exploring} found that LLMs could replicate most of the student-generated requirements, but emphasised limitations related to the quality of the prompts without addressing ethical considerations. Increased engagement and creativity were highlighted in \cite{mellqvistexploring}. They advocate for a hybrid AI-human approach, cautioning against overreliance and trust issues. A conceptual roadmap for utilising ChatGPT in Agile REE is presented in \cite{abdelfattah2023roadmap}, suggesting its transformative potential, but without empirical validation of its effectiveness with students. Additionally,  Moravánszky \cite{moravanszky2024banning} provided an experience report stressing the importance of prior RE knowledge in effectively guiding AI tools. }
\changed{In contrast to related studies, our study provides a large sample of students' views from two universities. To the best of our knowledge, this is the largest study on students' perceptions of LLM-supported REE to date. Our research presents an empirically based contribution to the field of REE, exploring learning outcomes, motivation, teamwork dynamics, and ethical considerations. In particular, it reveals how the effectiveness of a guided integration varies between individual and team-based coursework, aspects often overlooked in previous research.}

\subsection{Practical implications for REE}
While LLMs can enhance RE practices, Wiegers and Beatty \cite{wiegers2013software} caution that RE demands critical thinking and trade-off analysis. This study sheds light on how LLMs can support REE, but also draws attention to associated challenges, particularly those like ethical considerations and the risk of overreliance on AI-generated content. Several students expressed concern that such dependence could adversely affect their future decision-making and critical thinking skills.

To enhance the pedagogical value of LLMs in REE, coursework should integrate both human-led and AI-assisted analysis. Our findings indicate that students perceived the most effective approach as first completing RE tasks independently, followed by using LLMs for refinement and validation. This structured sequencing supports the development of critical thinking, deeper conceptual understanding, and analytical reasoning, ensuring that LLMs function as cognitive enhancers rather than substitutes. Furthermore, early structured training on the capabilities and limitations of LLMs is essential, particularly for novice users. Such training should include prompt engineering techniques and critical evaluation of LLM-generated content. By embedding these practices into assignments, educators can maximise the benefits of LLMs while mitigating some of their associated risks, thereby promoting a balanced and responsible use of LLMs into REE. 

The most commonly perceived challenges in integrating LLMs into coursework were related to academic integrity. Students expressed concerns about the ethical and responsible use of LLMs, the appropriate balance between AI-generated content and human input, and the proper referencing of LLM-generated contributions. This highlights the need to define clear ethical guidelines for using LLMs in coursework to help address students' concerns.

Further research is needed to examine the role of LLMs in supporting RE tasks within a team-based context and their impact on students' performance. Future studies should explore how LLMs can be effectively adapted to support structured RE processes and enhance collaboration in Agile RE environments, while ensuring that human critical thinking and teamwork remain central to the learning experience.

\section{Threats to Validity}
We adopted the threats to validity framework recommended by Wohlin et al.\cite{wohlin2012experimentation} to ensure a rigorous analysis. Like in every empirical study, potential limitations were considered.

\textit{Internal validity} threats arise from the reliance on self-reported survey data, which may introduce biases such as misinterpretation. To mitigate this, we analysed response distributions to identify trends and improve reliability while anonymising the survey to reduce overly positive responses. Although bonus point incentives raised concerns about rushed or overly favourable responses, the longer-than-expected average completion time, i.e. 15 minutes and 10 seconds, implies that students responded thoughtfully. Moreover, the course grading structure ensured that participation remained entirely voluntary, with no penalties for opting out, thereby reducing pressure to provide biased responses.  

\textit{External validity} is a potential concern as the study was conducted at only two universities, which may limit the generalizability of the findings. However, both institutions were selected because their RE courses reflect commonly used instructional strategies in REE, i.e. one emphasising individual assignments and blended learning, and the other leveraging team-based Agile projects. While this provides meaningful insights into different pedagogical approaches, future research should expand the sample to include universities with varying curriculum structures to validate broader applicability. In addition, the high response rate provides valuable insights into students' perceptions of AI-assisted learning, serving as a helpful reference for educators integrating LLMs into coursework. 

\textit{Construct validity} may be affected by variations in how students interpreted survey questions\changed{, which may also inherently affect \textit{conclusion validity}}. To address this, we included closed- and open-ended questions, which showed strong response alignment. \changed{A further limitation was the practical impossibility of enforcing the instructed use of LLMs. Although students were instructed to conduct human-led analysis before using LLMs for refinement and compare their outputs with those generated by the LLMs, ensuring complete adherence to these guidelines during independent assignment completion was practically impossible. This uncontrolled variable may have affected the authenticity of student reflections and the validity of perceived benefits, as the human-led nature of the initial analysis could not be reliably verified.} As part of future studies, we will incorporate student performance metrics, such as assignment grades and exam results, to assess the impact of LLMs on learning outcomes, critical thinking, and problem-solving. This will strengthen \textit{conclusion validity} by providing objective performance measures to complement self-reported perceptions.

%% file: sections/conclusion.tex
\section{Conclusion}
Integrating LLMs into REE has demonstrated substantial benefits and notable challenges. This study, conducted in UA and UB, highlights that LLMs significantly enhance student engagement, motivation, and understanding of RE concepts. LLMs can be valuable tools to support RE learning and professional practices by providing real-time assistance with requirements elicitation, documentation, stakeholder analysis, and prioritisation. Students generally perceived LLMs as useful for brainstorming, improving efficiency, and reducing time spent on repetitive tasks. However, concerns such as overreliance on AI, the accuracy of the responses generated, and how to integrate them properly in assignments were prevalent among students. The ethical implications of using LLMs represented the most significant challenge for students. Our study also found that the modality in which LLMs were integrated into assignments (individual vs. team-based) potentially influenced students' perceptions of the effectiveness of LLMs. UA students, who used LLMs primarily for individual assignments, reported more positive experiences. In contrast, UB students who worked on team-based project assignments experienced additional challenges in integrating AI-generated outputs into collaborative tasks. 
Despite these challenges, most students expressed satisfaction with the integration of LLMs in the course, and more than 75\% recommended the use of LLMs in future RE courses. The findings suggest that, while the guided use of LLMs offers significant support to the RE learning process, more structured guidelines and critical evaluation of LLM-generated content are essential to maximise benefits while preserving analytical thinking and academic integrity. Furthermore, for all students to experience the benefits of LLMs in RE practices, students with little or no prior experience with LLMs should receive additional guidance and resources on how to use them effectively to support learning. 

Future research should aim to refine the strategies for the guided use of LLMs, to improve AI literacy training for students, and to develop hybrid models that combine human expertise with AI-assisted support. It should also investigate how AI tools can facilitate collaboration in ASD teams. Moreover, developing strategies to assess the impact of LLM integration on student performance is essential to substantiate its educational value further.
Addressing these aspects can ensure an effective, balanced, and ethically responsible adoption of these learning tools.